# Temperature dependent motion of a massive quantum particle

Jian-Ping Peng

Department of Physics, Shanghai Jiao Tong University, Shanghai 200240, China

**Abstract**

We report model calculations of the time-dependent internal energy and entropy for a single quasi-free massive quantum particle at a constant temperature. We show that the whole process started from a fully coherent quantum state to thermodynamic equilibrium can be understood, based on statistics of diffracted matter waves. As a result of thermal interaction between the particle and its surroundings, the motion of the particle shows new feature.

# 1. Introduction

Recent experiments of absorption of a single photon by a single atom[1], nondestructive cooling of a single ion using an ultracold atomic bath[2], and the imaging of spin direction of a single atom[3] have demonstrated intriguing possibilities of controlling and measuring of quantum processes on the atomic scale. Among couplings of such a system with its surrounding, keeping it at adequate low temperature is believed to be essential to maintaining the quantum coherence. The question appears to be important and interesting: To what extent can temperature affect the behavior of a quantum system containing a single or a few quantum particles? It is well-known that statistical mechanics deals with macroscopic systems consisting of large number of microscopic particles[4]. For a system of $N$ particles at a finite temperature, it is generally true that the relative fluctuations of an extensive quantity vary as $1/\sqrt{N}$, which gets larger and larger as N decreases. If we wish to keep the statistical error below one percent then a system would have to contain more than about ten thousand particles. The obstacle arises if one tries to apply the standard procedure of statistical physics in a present-day textbook directly to study the temperature dependent behavior of a quantum system containing a few quantum particles. For an atom confined in a harmonic potential, it could be overcome by considering a single particle Hamiltonian coupled to a multi-mode quantum thermal bath, where relaxation of the particle depends on the coupling modes with acceptable fluctuations[5, 6]. Another approach is to perform statistics directly on the matter wave of a quantum particle by taking into account the detailed configuration of diffraction in real space, where the bath acts as the heat reservoir at a constant temperature and is large enough to cause unacceptable errors[7]. Here we improve the latter approach and study the thermodynamic process of a quasi-free massive quantum particle from a coherent single quantum state to thermodynamic equilibrium. We show analytical expressions of the time-dependent internal energy and entropy. Due to thermal interaction between the particle and its surroundings, the process of decoherence is temperature dependent and its translational motion shows new feature.

**2. Model Calculations and Results**

We consider a structureless massive quantum particle moving in a space at constant temperature $T$ as shown in Fig. 1. Although there is no interaction with other particles, the particle is quasi-free because a circular aperture of radius $a_0$ for its matter wave is considered. The quantum particle is assumed to be of kinetic energy $E_0$ initially at the origin and is described by a wave-packet sharply peaked at the de Broglie wavelength $\lambda = h/\sqrt{2mE_0}$, with $m$ being its mass and $h$ the Plank's constant. The wave front propagates along the x-axis at group velocity $V_g = h/(m\lambda)$. For simplicity, we assume that the radius $a_0$ of the matter wave aperture is large compared with the wavelength $\lambda$ so that the shape and linear dimension of the forward-going wave-front remains unchanged as the wave propagates. It is important that every point at the edge of the wave-front generates out-going fully spherical waves and thus the kinetic energy associated with the forward-going wave-front follows the form[7]

$$E_k(x) = E_0 \exp(-2\lambda x / a_0 L) \quad , \tag{1}$$

where $L$ is a temperature dependent parameter of dimension length and is expected to be infinitely large as the temperature tends to zero. This is just the energy for the source to generate out-going fully spherical waves. We define a probability density function for the particle

$$P_E(x) = (2\lambda / a_0 L) \exp(-2\lambda x / a_0 L). \tag{2}$$

There exists a step length $d_0$, so that the energy for fully spherical waves generated in $d_0$ is $E_0 P_E(nd_0)d_0$ when the forward-going plane-wave front is at the position $nd_0$, where $n$ is an integer. In principle, the particle may be in any of these diffracted states besides the forward-going plane-wave state, i.e., the particle itself constitutes automatically a thermodynamic system as a result of diffraction at the edge of its matter-wave front. The probability for different energy states is known here and, in general, all energy states are not equally likely. If one can distinguish energy states in more detail, a smaller step $\Delta x$ should be used and the degeneracy replaced by $d_0/\Delta x$. Thermal interaction between the particle's system and the surrounding space becomes possible and the space here acts as the heat reservoir at constant temperature $T$.

The partition function of the particle should take into account contributions from forward-going wave-front and all spherical waves diffracted at the edge. The partition function at a given time is then

$$Z(t) = \int_0^{V_g t} P_E(x) \exp[-\beta E_0 d_0 P_E(x)] dx + \exp(-2\lambda V_g t/a_0 L) \exp[-\beta E_0 \exp(-2\lambda V_g t/a_0 L)]$$

$$= [e^{-Z_0 \exp(-t_r)} - e^{-Z_0}]/Z_0 + \exp(-\beta E_0 e^{-t_r} - t_r), \qquad (3)$$

where the notation $\beta = 1/k_B T$ is used as usual with $k_B$ being the Boltzmann constant. The constant $Z_0$ is defined as the non-zero real solution of the transcend equation

$$\exp(Z_0) - 2Z_0 - 1 = 0 \qquad (4)$$

With the help of the Lambert W function[8], we obtain $Z_0 = -W_{-1}(1/2e^{1/2}) - 1/2$ and we use an approximate value $Z_0 = 1.25643$ in our numerical calculations. Here we have used the correct form of the step length $d_0 = Z_0 a_0 L/(2\lambda \beta E_0)$ instead of the wavelength, and $t_r = t/t_c$ is the time scaled with a temperature dependent characteristic time defined by

$$t_c = m a_0 L / 2h. \qquad (5)$$

The expectation value for the energy of the particle or its internal energy is

$$U(t) = \frac{1}{Z(t)} \int_0^{V_g t} E_0 d_0 P_E^2(x) \exp[-\beta E_0 d_0 P_E(x)] dx$$

$$+ \frac{E_0}{Z(t)} \exp(-4\lambda V_g t / a_0 L) \exp[-\beta E_0 \exp(-2\lambda V_g t / a_0 L)]. \quad (6)$$

The first term comes from all those out-going spherical waves diffracted at the edge, and the second term from the forward-going wave-front itself as a whole. The internal energy of the particle is written as

$$U(t) = \frac{N_1 + N_2}{D} \times \frac{1}{\beta}, \quad (7)$$

with

$$N_1 = Z_0 [\exp(\beta E_0 e^{-t_r} + Z_0) - \exp(\beta E_0 e^{-t_r} + t_r + Z_0 e^{-t_r})],$$

$$N_2 = \beta E_0 Z_0 \exp(Z_0 + Z_0 e^{-t_r} - t_r) + [e^{Z_0 + t_r} - \exp(Z_0 e^{-t_r} + t_r)] \exp(\beta E_0 e^{-t_r}),$$

$$D = [e^{Z_0 + t_r} - \exp(Z_0 e^{-t_r} + t_r)] \exp(\beta E_0 e^{-t_r}) + Z_0 \exp(Z_0 + Z_0 e^{-t_r}).$$

It depends on the temperature and the particle's initial energy in a complicated form.

In Fig. 2 we plot the numerical results of the internal energy for various initial energies and temperatures. In general, a quantum particle absorbs or gives out heat continuously when its initial energy $E_0$ is less or more than $k_B T/2$. At the special point $E_0 = k_B T/2$, our numerical calculations show that the curve moves downward initially and then upward continuously, indicating exchange of energy taking place between the particle and its surrounding space in the whole process. The limiting value of the internal energy for the freedom in the x-direction is $k_B T/2$, regardless of its initial energy. The limit is already reached for $t/t_c > 3$ in our numerical calculations and the overall decay or increase in internal energy does not follow the simple exponential form. Physically, the process describes the evolution of the particle starting from a fully coherent initial state with well defined energy in quantum mechanics to a series of possible states. Although states representing the forward-going wave-front remain coherent, those states diffracted at the edge at different time are no longer coherent at finite temperatures as a result of heat exchange. The process breaks down if a collision occurs in a gas. After a sufficient long time, the probability for the particle in the coherent state disappears and the internal energy reaches the universal thermodynamic energy in equilibrium with the surrounding environment. The time needed to reach the limit of internal energy could in some sense be viewed as the coherence decay time of a quantum particle in a space at a finite temperature. Therefore, the present work might be utilized in understanding the temperature dependence for the coherence decay time observed recently for individual molecules at room temperature[9].

We now calculate the average value of the particle's position to obtain information of where it is at a given time. Although out-going spherical waves diffracted at the edge are no longer coherent at finite temperatures, the overall probability projected on $x$-axis in the range $-V_g t < x < V_g t$ can be calculated without mathematical difficulties. The thermodynamic expectation of the particle's position in the $x$-direction is written as

$$\bar{x}(t) = \frac{1}{Z(t)} \int_{-V_g t}^{V_g t} x \cdot \{ \int_0^{(V_g t + x)/2} \frac{1}{2} \frac{P_E(x_1) \exp[-\beta E_0 d_0 P_E(x_1)]}{V_g t - x_1} dx_1 \} dx$$

$$+ \frac{V_g t}{Z(t)} \exp(-2\lambda V_g t / a_0 L) \exp[-\beta E_0 \exp(-2\lambda V_g t / a_0 L)]$$

$$= \frac{1}{Z(t)} \{ \int_0^{V_g t} x P_E(x) \exp[-\beta E_0 d_0 P_E(x)] dx + \frac{a_0 L}{2\lambda} t_r \exp[-t_r - \beta E_0 \exp(-t_r)] \} . \qquad (8)$$

The first term represents contributions from all those out-going spherical waves and the second term from the forward-going wave-front. The explicit result is

$$\bar{x}(t) = \frac{[Ei(-Z_0 e^{-t_r}) - Ei(-Z_0)] \exp[(\beta E_0 + t_r e^{t_r} + Z_0) e^{-t_r}]}{\exp(\beta E_0 e^{-t_r} + t_r) + Z_0 \exp(Z_0 e^{-t_r}) - \exp(\beta E_0 e^{-t_r} + Z_0 e^{-t_r} + t_r - Z_0)} \frac{a_0 L}{2\lambda}$$

$$+ \frac{Z_0 t_r \exp(Z_0 e^{-t_r}) + t_r \exp(\beta E_0 e^{-t_r} + t_r)}{\exp(\beta E_0 e^{-t_r} + t_r) + Z_0 \exp(Z_0 e^{-t_r}) - \exp(\beta E_0 e^{-t_r} + Z_0 e^{-t_r} + t_r - Z_0)} \frac{a_0 L}{2\lambda}, \quad (9)$$

with $Ei(z)$ being the exponential integral function of argument $z$[10]. Note that the particle, in principle, may be found anywhere in the range $-V_g t < x < V_g t$. It is interesting that $\bar{x}(t)$ shows dependences not only on the initial energy of the particle, but also on the temperature and the spatial confinement for its matter wave.

Fig. 3 shows the numerical results of Eq. (9) for the scaled average position $\bar{x}(t)/(a_0 L/\lambda)$ versus the scaled time $t/t_c$ for various combinations of temperature and initial energy. It increases continuously with increasing time, approaching a limit within a time of several $t_c$. It is true that a particle with higher initial energy reaches the limit at a later time. The limit is

$$\bar{x}(\infty) = \frac{a_0 L}{2\lambda} \frac{1}{Z(\infty)} \times \int_1^0 e^{-Z_0 l} \ln l \, dl$$

$$= \frac{C + \ln Z_0 - Ei(-Z_0)}{1 - e^{-Z_0}} \times \frac{a_0 L}{2\lambda} \approx 1.33 \times \frac{a_0 L}{2\lambda}, \quad (10)$$

where $C=0.577216$ is the Euler's constant[10]. If the particle could not feel the temperature of the space, we would have the relation $\bar{x}(\infty) = a_0 L/2\lambda$.

The entropy, measuring the amount of uncertainty of the particle, can also be calculated because the probability distribution is known. The uncertainty relation $\Delta E \cdot \Delta t \approx h/2\pi$ in quantum mechanics tells us that waves diffracted at the edge should be indistinguishable when the wave-front moves forward within about one wavelength. We choose $\lambda$ as the step length and rewrite the probability in discrete form $\sum p_i = 1$. The entropy is calculated using the definition $S_x = -k_B \sum p_i \ln p_i$, which is perfectly unambiguous for time-dependent system of any size. Under the condition $\lambda \ll V_g t_c$, we find the time-dependent entropy of the particle for the coordinate is

$$S_x(t) = \frac{U(t)}{T} + k_B \ln Z(t) + k_B \frac{2\lambda}{a_0 L} \bar{x}(t)$$

$$= \frac{U(t)}{T} + k_B \ln Z(t) + \frac{k_B}{Z(t)} \{ \int_1^{\exp(-t_r)} e^{-Z_0 l} \ln l \, dl + t_r \exp[-t_r - \beta E_0 \exp(-t_r)] \}. \tag{11}$$

Here a term of order $(\lambda/V_g t_c)\ln(\lambda/V_g t_c)$ has been neglected. The last term can be understood as the result of expansion of the particle's matter-wave. The entropy starts from zero and then increase monotonically with increasing time. As $U(t)$ and $\bar{x}(t)$ reach their limits, the entropy for each freedom converges to a universal value

$$S_x(\infty) = k_B [\frac{1}{2} + \ln \frac{1-e^{-Z_0}}{Z_0} + \frac{C + \ln Z_0 - Ei(-Z_0)}{1 - e^{-Z_0}}] \approx 1.27 k_B, \tag{12}$$

which is independent of the temperature and the initial state of the particle. The result holds also for a free quantum particle. As a consequence of the wavelike nature of the particle, the feature is purely quantum mechanical without classical counterpart.

The possibility of experimental observation depends on developing single quantum particle experiment in which there is very little chance for two wave packets to overlap. It seems to have been demonstrated by past experiments on the diffractions of neutrons[11], electrons[12] and atoms[13]. For electrons, the length of a single electron wave packet could be as short as 1 μm[12]. It is difficult at present to observe directly the instantaneous position of a single quantum particle, since an unavoidable quantum coupling with the apparatus must be invoked in the measurement[14,15]. However, it is possible to experiment with a beam of incident particles at a single wave-length. If the final positions of the incident particles are recorded by a row of detectors on *x*-axis, the average value of position predicted in Eq. (10) should be observable. An alternative way is to measure the portion of particles reached on the target at a reasonable distance apart at various temperatures. Observation of the dependences on temperature and the distance would be a unique demonstration of the response of a single quantum particle to temperature of the space around it.

## 3. Conclusions

In conclusion, the motion of a quasi-free massive quantum particle in a space at a finite temperature shows new phenomena arising from thermal interaction between them. The important point is that a nonzero temperature of the environment inherently destroys the coherence of a quantum particle and must be taken into account in future device development based on a single or a few of quantum mechanical states. Applications of the present work might be found in the struggle against environmental decoherence of small quantum systems, in understanding the detailed temperature dependent behavior of ballistic transport in nanoscale structures, and even that of an atom in a gas between two successive atomic collisions.

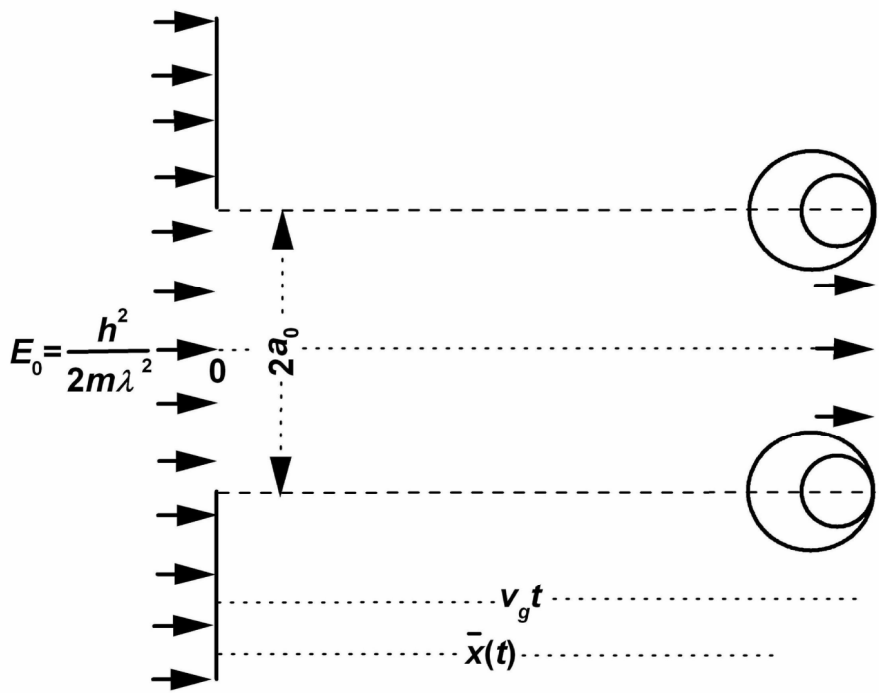

Figure 1. Cross section schematic of a free quantum particle passing through a matter wave aperture of radius $a_0 \gg \lambda$. The whole space is assumed to be at constant temperature $T$. Arrows in line indicate the wave-front of the mater wave pulse. Circles indicate out-going spherical waves diffracted at the edge of the wave-front. The expectation value $\bar{x}(t)$ for the position of the particle is temperature dependent.

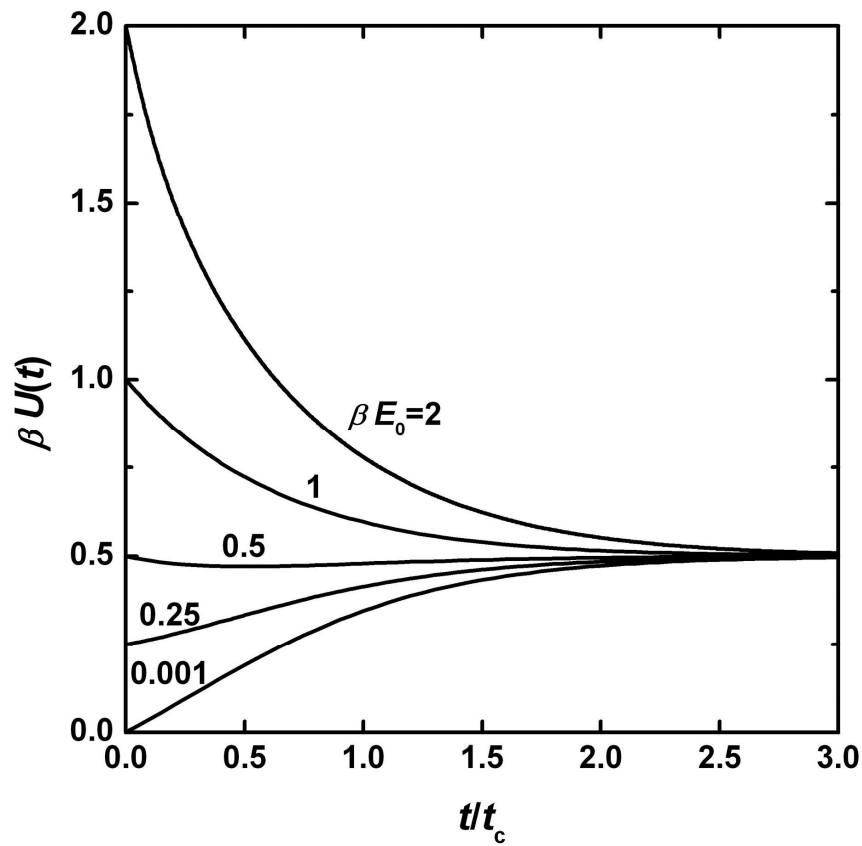

Figure 2. Scaled internal energy $\beta U(t)$ as a function of the scaled time $t/t_c$. Values of $\beta E_0$ are labeled on the curves.

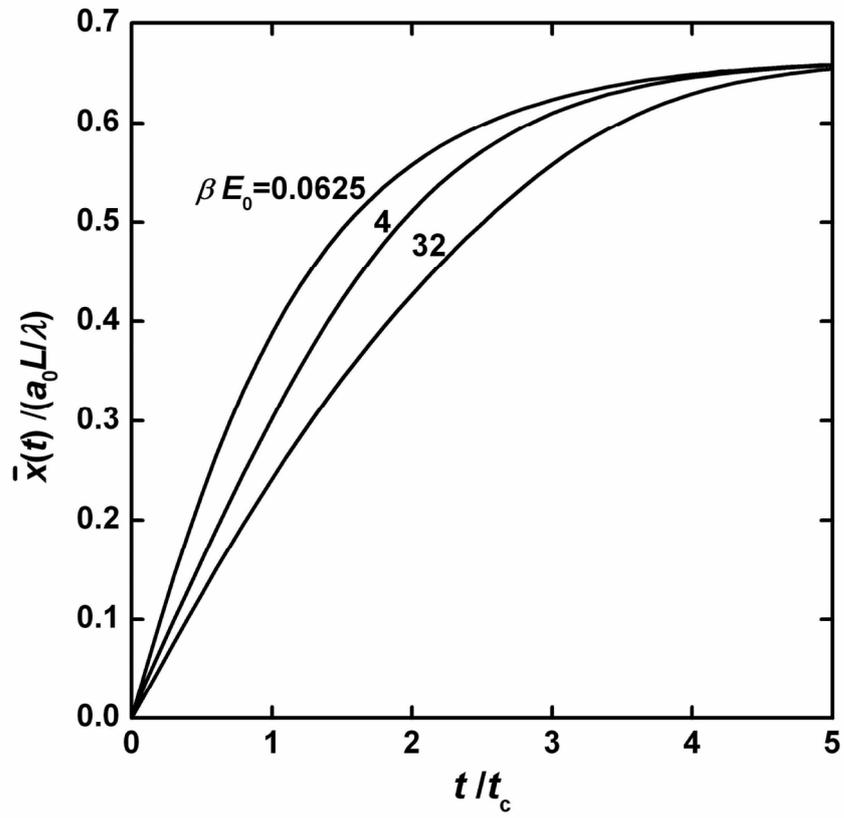

Figure 3. Scaled average position $\bar{x}(t)/(a_0 L/\lambda)$ versus the scaled time $t/t_c$ for various values of $\beta E_0$.